\documentclass[twocolumn]{aastex631}

\usepackage{newtxtext,newtxmath}
\usepackage{wrapfig}

\begin{document}

\shorttitle{Enhancing MHD model accuracy and CME forecasting with Faraday Rotation}
\shortauthors{Salvatore Mancuso}

\newcommand{\kms}{km~s$^{-1}$}
\newcommand{\mss}{m~s$^{-2}$}
\newcommand{\dg}{$^{\circ}$}
\newcommand{\rsun}{$R_{\odot}$}
\newcommand{\Alfven}{Alfv\'en}
\newcommand{\Alfvenic}{Alfv\'enic}

\title{\large Enhancing MHD model accuracy and CME forecasting \\ by constraining coronal plasma properties with Faraday rotation}

\author{\large Salvatore Mancuso}
\affiliation{Istituto Nazionale di Astrofisica, Osservatorio Astrofisico di Torino, via Osservatorio 20, Pino Torinese 10025, Italy}


\begin{abstract}
Accurate forecasting and modeling of coronal mass ejections (CMEs) and their associated shocks are pivotal for understanding space weather and its impact on Earth. 
This requires a detailed understanding of CMEs' 3D morphology and the properties of the pre-eruption coronal plasma, which are usually inferred from global 3D numerical magnetohydrodynamic (MHD) simulations.
Refining MHD models is thus crucial for improving our understanding of CME-driven shocks and their effects on space weather.
Faraday rotation measurements of extragalactic radio sources occulted by the solar corona serve as a powerful complementary tool for probing the pre-eruption electron density and magnetic field structure. 
These measurements thereby allow us to refine predictions from global MHD models.
In this paper, we discuss our recent study of the morphological evolution of a CME-driven shock event that occurred on August 3, 2012. 
Our analysis used white-light coronagraphic observations from three different vantage points in space: the Solar and Heliospheric Observatory (SOHO) and the twin Solar Terrestrial Relations Observatory (STEREO) spacecraft A and B. 
Obtaining data from these spacecraft, we derived key parameters such as the radius of curvature of the driving flux rope, the shock speed, and the standoff distance from the CMEs' leading edge.
A notable feature of this event was the availability of rare Faraday rotation measurements of a group of extragalactic radio sources occulted by the solar corona, which were obtained a few hours before the eruption.
These observations from the Very Large Array (VLA) radio interferometer provide independent information on the integrated product of the line-of-sight (LOS) magnetic field component and electron density.
By modeling the shock standoff distance and using constraints from the Faraday rotation measurements, we achieve a high level of agreement between the fast-mode Mach number predicted by the Magnetohydrodynamic Algorithm outside a Sphere (MAS) code in its thermodynamic mode and the value deduced from the analysis of the 3D reconstruction of coronagraphic data, provided that appropriate correction factors ($f_b \simeq 2.4$ and $f_n \simeq 0.5$) are applied in advance to scale the simulated magnetic field and electron density, respectively. 
Our results are consistent with previous estimates and provide critical information for fine-tuning future MHD simulations.
\end{abstract}



\section{Introduction}

Understanding the physical properties of the solar coronal plasma is critical for uncovering the mechanisms underlying a vast number of transient phenomena in space plasma physics and for forecasting space weather with a greater accuracy. 
Determining the properties of violent, large expulsions of plasma and magnetic flux known as coronal mass ejections (CMEs) before they impact Earth is crucial. 
This is because interplanetary CMEs, which propagate from the Sun to Earth in only a few days, typically cause strong space weather disturbances in the near-Earth space environment.
When an impulsively accelerated magnetized plasma expands through the corona at a velocity exceeding the local characteristic speed, the outer boundary of its expanding ejecta forms a large-amplitude fast-mode magnetohydrodynamic (MHD) wave. 
Nonlinear effects can cause this wave to steepen into a shock. 
Both the shock and the ejecta contribute to geomagnetic disturbances and are significant accelerators of solar energetic particles (SEPs). 
Thus, understanding their independent evolution is crucial for making accurate space weather predictions.

Although global MHD models of the solar corona and solar wind have become increasingly sophisticated, our ability to predict the space-weather impact of CMEs and their associated shocks remains limited. 
The primary sources of uncertainty in predicting a CME's arrival time at Earth include the initial properties of the ejecta--its speed, mass, and propagation direction--as well as the characteristics of the ambient solar wind through which it propagates (\citealt{Riley2021a,verbeke2023}). 
These MHD models are particularly sensitive to the pre-CME distribution and magnitude of coronal magnetic fields and plasma electron densities. 
Despite recent progress, significant discrepancies remain between the predicted and observed magnetic field strengths (\citealt{riley2021b}). 
In fact, MHD models of the corona often underestimate the magnitude of the open magnetic flux when compared to in situ measurements (\citealt{linker2017}) and overestimate the solar electron density (\citealt{rouillard2016, kouloumvakos2019}).
In the absence of in situ measurements--such as those recently provided by NASA's Parker Solar Probe (PSP), launched in 2018--obtaining this information proves challenging. 

Faraday rotation observations of linearly polarized extragalactic radio sources occulted by the solar corona and wind are a powerful method for constraining coronal density and magnetic field strengths. 
This technique measures the integrated product of the line-of-sight (LOS) component of the magnetic field $B_\parallel$ and the electron density $n_e$, which yield critical information about the plasma environment through which CMEs and shocks propagate. 
Faraday rotation relies on the physical effect in which a linearly polarized radio wave propagating through a plasma undergoes a rotation of its plane of polarization. 
The induced rotation of the radio wave's polarization plane is defined by $\Delta\chi=\lambda^2$RM, where $\lambda$ is the wavelength of the radio signal and RM is the rotation measure, given by
\begin{equation}
\mathrm{RM}={e^3\over{2\pi{m_e}^2c^4}}\int_{\rm LOS}{n_e}{\bf B}\cdot{\bf ds}.
\end{equation}
In this equation, \( n_e \) is the electron density; \( \mathbf{B} \) is the vector magnetic field; \( \mathbf{ds} \) is the vector differential path length along the LOS; $c$ is the speed of light; and $e$ and $m_e$ are the charge and mass of the electron, respectively. 
The contribution to the RM is positive for a magnetic field pointed toward the observer. 
In past decades, when using observations from the Very Large Array (VLA) radio interferometer of the National Radio Astronomy Observatory (NRAO) at GHz frequencies, this technique has provided valuable information on coronal plasma and solar wind structures (\citealt{sakurai1994,mancuso1999,mancuso2000, spangler2000,spangler2005,ingleby2007,mancuso2013a,mancuso2013b,kooi2014,lechat2014,kooi2017}).

In our study, we used rare Faraday rotation measurements of a group of extragalactic radio sources that were occulted by the solar corona a few hours before the August 3, 2012, CME-driven shock event. 
These measurements allowed us to obtain the integrated product of the LOS component of the magnetic field and the electron density of the pre-event corona.
By integrating these data into the Magnetohydrodynamics Around a Sphere Thermodynamic (MAST) model, and modeling the shock standoff distance to estimate the fast-mode shock speed, we lastly demonstrate how correction factors may be determined and applied to accurately scale the simulated magnetic field and electron density for that period.
The paper is organized as follows: Section 2 presents an overview of the observations and reconstruction methods. Section 3 describes the MHD model and Faraday rotation observations. Section 4 provides our results. Section 5 outlines a summary and conclusion.

\begin{figure}
\centering
\includegraphics[width=8.5cm]{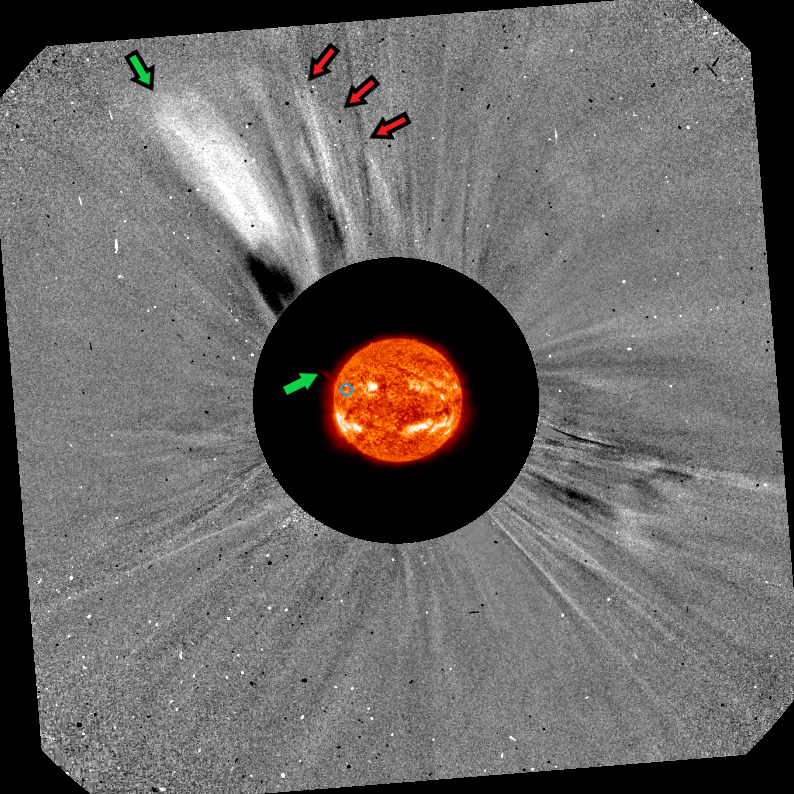}
\caption{
Composite image of the solar corona on August 3, 2012, at 04:17 UT spanning from the solar disk center to the extended corona up to approximately 6 \rsun (created using JHelioviewer; \citealt{muller2017}).
The running difference image from the SOHO/LASCO C2 white-light coronagraph shows a narrow CME ({\it green arrow}) propagating northeast, accompanied by the boundary of the spheroidal shock wave driven by the CME ({\it red arrows}). 
The inner image captured in the 304 \(\text{Å}\) passband by the SDO/AIA instrument depicts the earlier evolution of the CME. The {\it blue circle} marks the origin of the jetlike eruption on the Sun's surface.}
\label{Fig1}
\end{figure}

\begin{figure}[hb!]
\centering
\includegraphics[width=8.5cm]{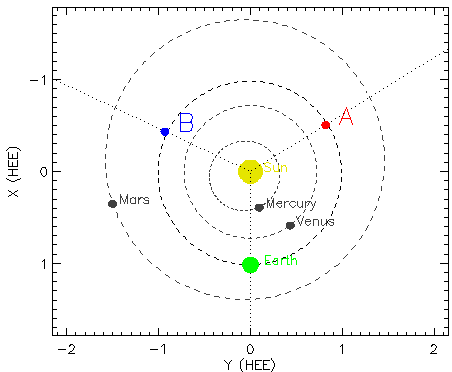}
\caption{Positions of the twin STEREO-A and STEREO-B spacecraft relative to the Earth on August 3, 2012, at 03:00 UT. 
The dotted lines show the angular displacement from the Sun. The heliocentric Earth ecliptic (HEE) distance is expressed in AU. Courtesy of http://stereo-ssc.nascom.nasa.gov/where.shtml.
}  
\label{Fig2}

\end{figure}
\begin{figure*}
\centering
\includegraphics[width=6cm]{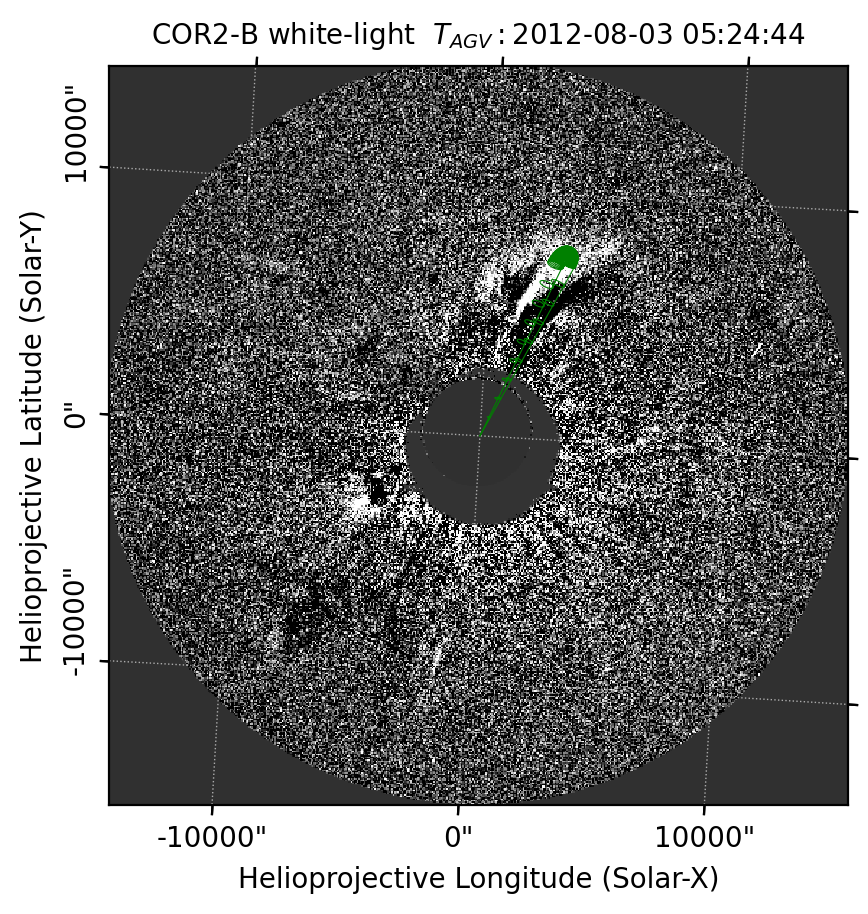}
\includegraphics[width=6cm]{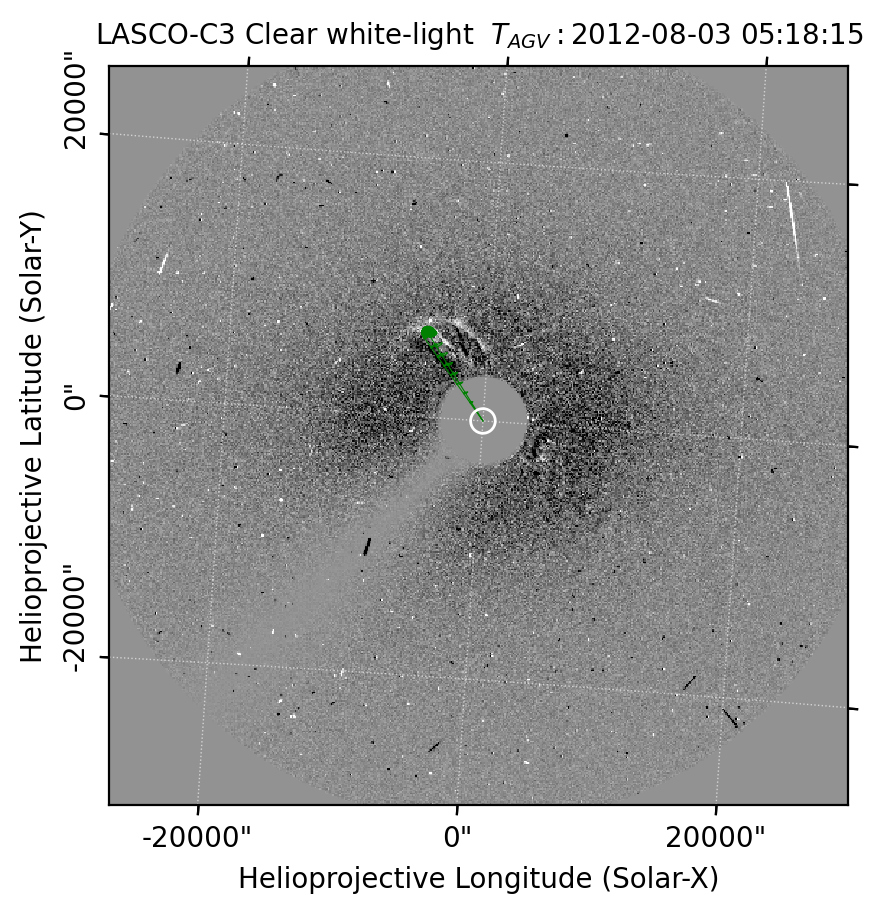}
\includegraphics[width=6cm]{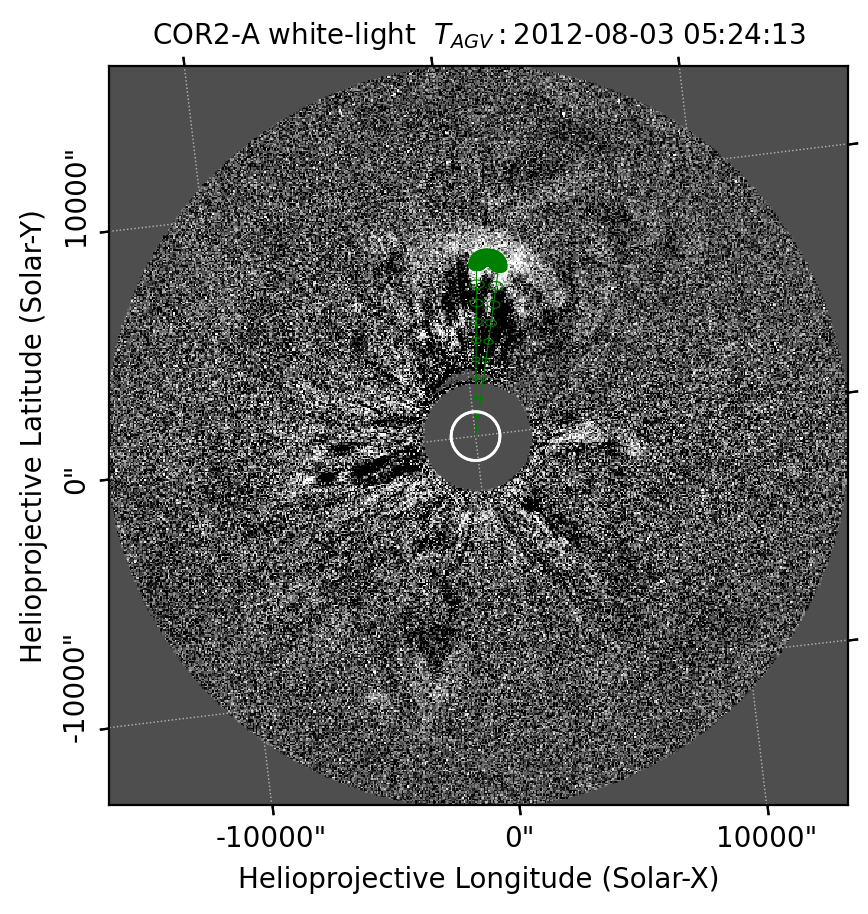}
\includegraphics[width=6cm]{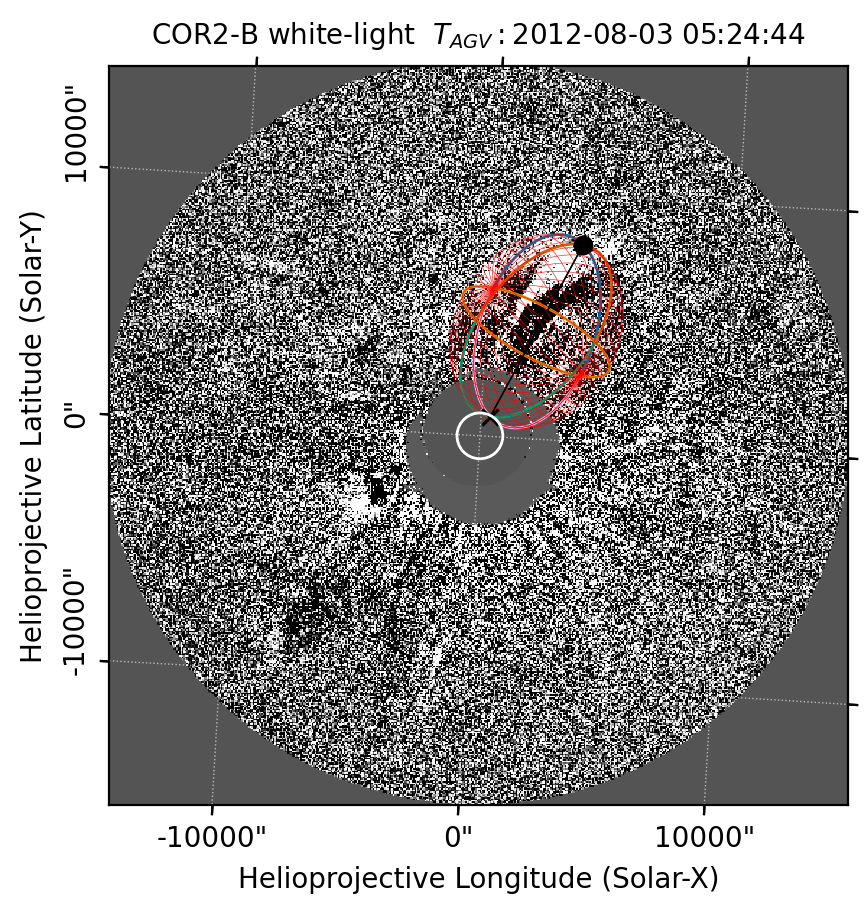}
\includegraphics[width=6cm]{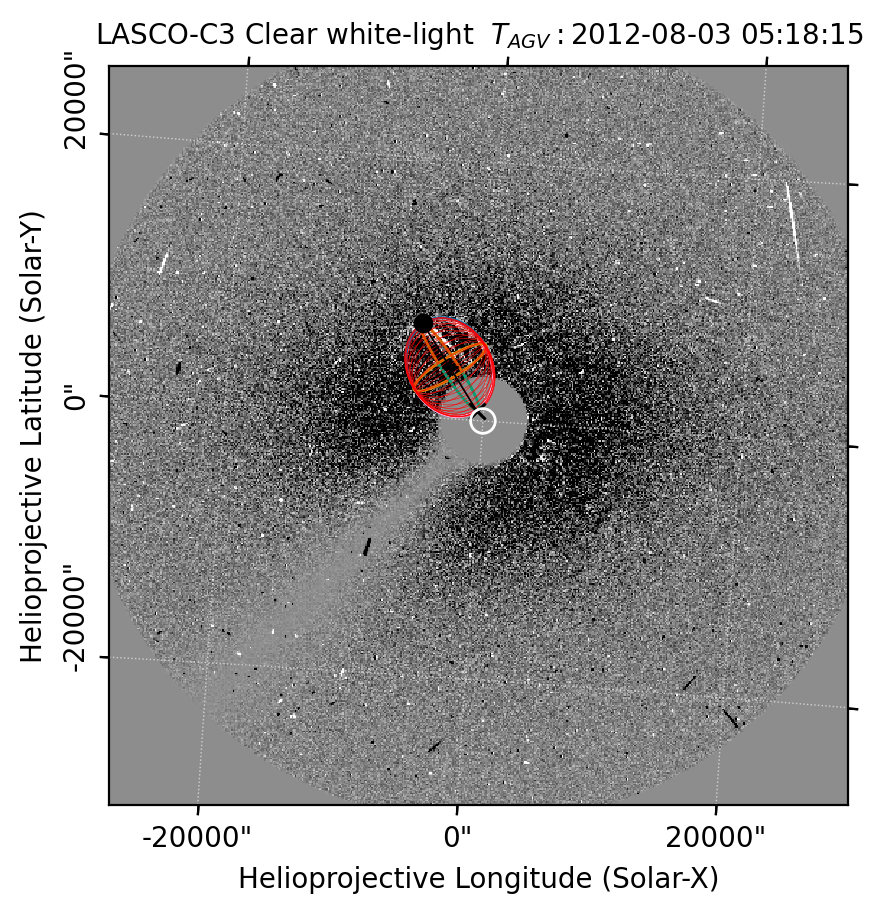}
\includegraphics[width=6cm]{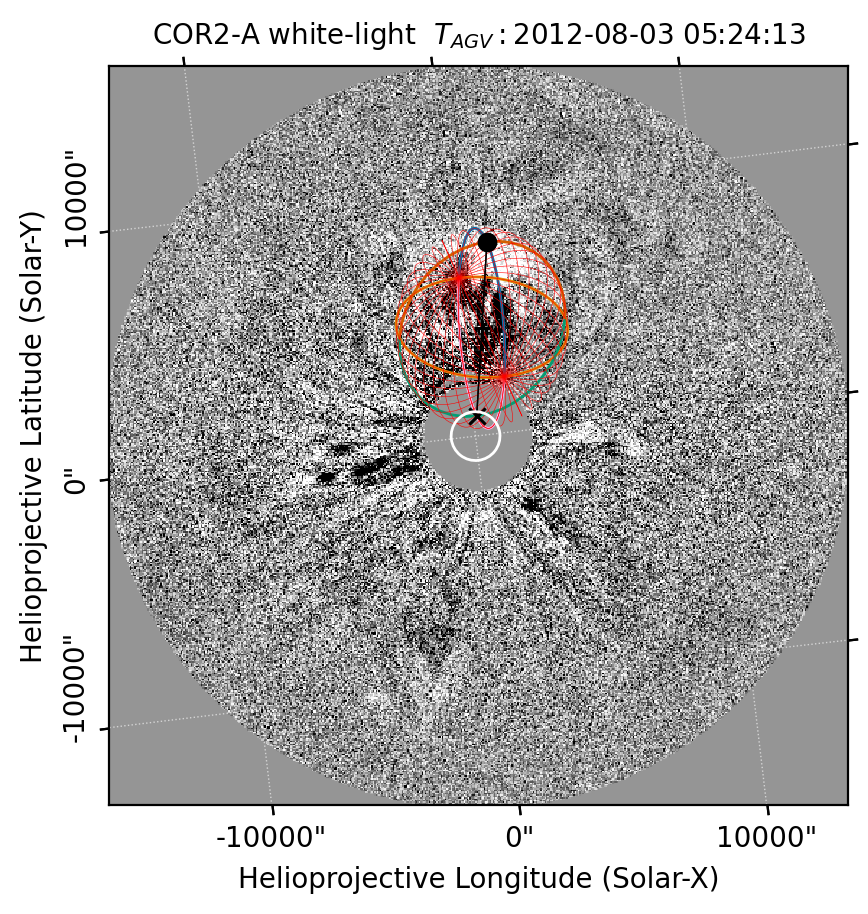}
\caption{
Running difference images from STEREO-B/COR2 ({\it left}), SOHO/LASCO/C3 ({\it middle}), and STEREO-A/COR2 ({\it right}) taken around the same time on August 3, 2012, showing the CME apex reaching a height of approximately 10 \rsun.
{\it Top panels:} GCS reconstruction of the flux rope shown in green. 
{\it Bottom panels:} Reconstruction of the shock surface depicted as a spheroidal surface superposed on the observed images.  
For further details, see the main text.}  
\label{Fig3}
\end{figure*}

\section{The August 3, 2012, CME-driven shock}

\subsection{Observations}

The August 3, 2012, CME and associated shock event occurred during the Carrington rotation (CR) number 2126. 
It coincided with a weak, impulsive B-class flare that peaked at approximately 2:33 UT above the National Oceanic and Atmospheric Administration Active Region (NOAA AR) 11537. 
At the time of the eruption, this region was located at N13 E57. 
Around 2:30 UT, a group of coronal type III radio bursts at meter wavelengths marked the impulsive phase of the flare. 
These bursts, which are radio signatures of suprathermal electron beams propagating along open magnetic field lines, are typically associated with magnetic reconnection events at the flare site (\citealt{reed2014}).

As observed by the Atmospheric Imaging Assembly (AIA; \citealt{Lemen2012}) instrument on board the Solar Dynamics Observatory (SDO; \citealt{pesnell2012}) in the 304 \AA\ passband (Fig.~\ref{Fig1}), the initial appearance of the CME was jetlike.
Farther up in the corona, a narrow CME surrounded by a diffuse halo was observed by the Large Angle and Spectrometric Coronagraph (LASCO; \citealt{brueckner1995}) using the C2 and C3 coronagraphs aboard the Solar and Heliospheric Observatory (SOHO; \citealt{domingo1995}).
The C2 coronagraph covers the region from 2 to 6 \rsun, while C3 extends from 3.7 to 30 \rsun. Together, they provide white-light images of Thomson-scattered radiation from the extended corona.
The CME event was listed as a partial halo in the SOHO LASCO CME catalog (CDAW; \citealt{yashiro2004,gopalswamy2009}) and was first observed at a height of 2.71 \rsun\ at 3:12 UT within the LASCO C2 field of view. 
A linear fit to the CME’s radial expansion indicated a projected speed of 520 \kms, while a quadratic fit suggested an initial speed of about 650 \kms, with a deceleration of -8.5 \mss.

The CME was also observed from two different viewpoints by the inner coronagraph COR1 (1.5 to 4 \rsun) and the outer coronagraph COR2 (2.5 to 15 \rsun), both part of the Sun-Earth Connections Coronal and Heliospheric Investigation (SECCHI; \citealt{howard2008}) suite on board the twin Solar Terrestrial Relations Observatory (STEREO; \citealt{kaiser2008}) spacecraft. 
These spacecraft, STEREO-A and STEREO-B, are positioned on heliocentric orbits ahead of and behind Earth, respectively. 
At the time of the eruption, STEREO-A and STEREO-B were located at approximately W122 and E115, respectively, relative to Earth (see Fig.~\ref{Fig2}), a configuration that was particularly favorable for a 3D reconstruction of the eruption.
Despite its initial jetlike appearance, the white-light images provided by the two STEREO coronagraphs in the extended corona show that the eruption exhibited a definite, albeit narrow, flux rope structure (see Fig.~\ref{Fig3}).
Furthermore, the difference images from the three coronagraphs clearly show a faint, diffuse spheroidal region in front of the flux rope, interpreted as the shock surface marking the boundary of the sheath region.

\subsection{CME and shock 3D reconstruction}

The eruption of CMEs is commonly attributed to the destabilization of coronal magnetic arcades, which contain semitoroidal magnetic flux ropes anchored in opposite photospheric magnetic polarities (e.g., \citealt{green2018}). 
Once the pre-eruptive structure loses equilibrium, it accelerates due to the unbalanced Lorentz force and expands in a self-similar manner (e.g., \citealt{Low1984, Thernisien2009}).
The bright flux rope-like structure of the August 3, 2012, CME was fitted using the graduated cylindrical shell (GCS; \citealt{Thernisien2009,thernisien2011}) model. 
This 3D empirical model of a magnetic flux rope represents the structure as a hollow, croissant-like shell. 
The GCS model is widely used to reconstruct the large-scale geometry of flux rope-like CMEs in the solar corona. 
The 3D reconstruction was performed using the Python-based software package PyThea (\citealt{kouloumvakos2022}). 
This involved overlaying a parameterized GCS model grid onto each image and adjusting its parameters until the fit closely matched the observations from the three spacecraft positioned along different LOSs. 
The flux rope's origin at a latitude of approximately 13\dg\ above the solar equator is indicated by the GCS reconstruction in the top panels of Fig.~\ref{Fig3}. 
Initially, the flux rope followed a nonradial trajectory, rapidly increasing its latitude by about 40\dg\ within an hour and reaching 5 \rsun\ by 4:00 UT. 
After 4:00 UT, its trajectory became nearly radial. 
This pattern agrees with previous results, which find most CME deflections and rotations to occur below approximately 5 \rsun\ (e.g., \citealt{kay2015}).
Due to uncertainties when modeling the initial phase of abrupt nonradial expansion--and because the Faraday RMs in this study refer to heliocentric distances greater than 8 \rsun, we limited our analysis to heights greater than 6 \rsun\ (i.e., where the trajectory became nearly radial).
The faint and diffuse spheroidal shock wave in front of the expanding flux rope was modeled as a prolate spheroid bubble (\citealt{kwon2014}), implying a fast-mode shock wave that expands self-similarly.
The bottom panel of Fig.~\ref{Fig3} shows a snapshot of the fits applied simultaneously to coronagraph observations from STEREO-A, STEREO-B, and SOHO. 

\begin{figure}
\centering
\includegraphics[width=8.5cm]{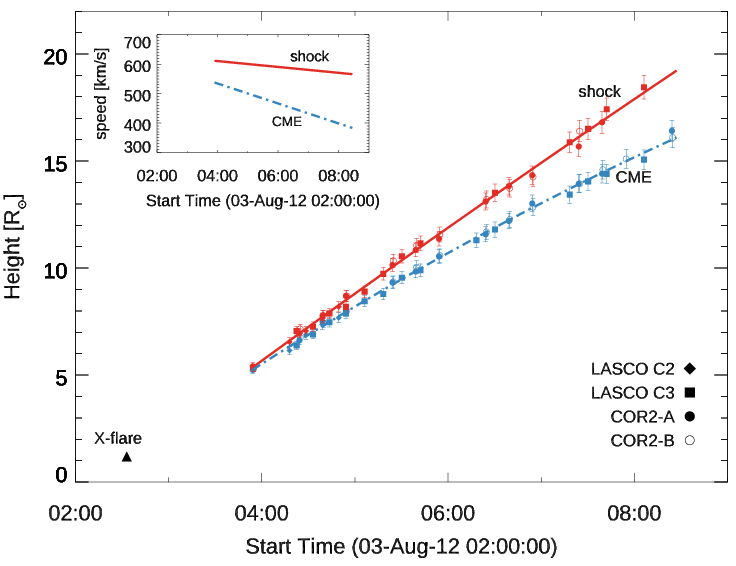}
\caption{Kinematic evolution of the leading edge of the CME flux rope and the nose of the associated shock, as derived from observations by SOHO, STEREO-A, and STEREO-B. The solid and dot-dashed lines represent second-order polynomial fits to the data for each front. The inset shows the inferred speed as a function of time for both fronts.}  
\label{Fig4}
\end{figure}

Figure~\ref{Fig4} shows the heliocentric distance-time profiles corresponding to the leading edge of the CME and the nose of the shock.
As indicated by the inset of Fig.~\ref{Fig4}, the leading edge of the flux rope and the shock decelerate as they move farther from the Sun due to drag forces acting on the CME. 
Magnetic tension and pressure gradients also contribute to the deceleration of erupting flux ropes (\citealt{lin2022}). 
The standoff distance \(\Delta R\) increases as the CME moves farther from the Sun because the shock speed exceeds the CME speed. 
Although driven bow shocks typically propagate at the same speed as their driver in homogeneous ambient media, this is not always the case. 
The anisotropic nature of the solar wind can lead to variations in behavior (\citealt{vrsnak2008}). 
This observation is consistent with 2.5D numerical MHD simulations of fast CMEs and their associated shock fronts (\citealt{savani2012}).

\begin{figure*}
\centering
\includegraphics[width=13.5cm]{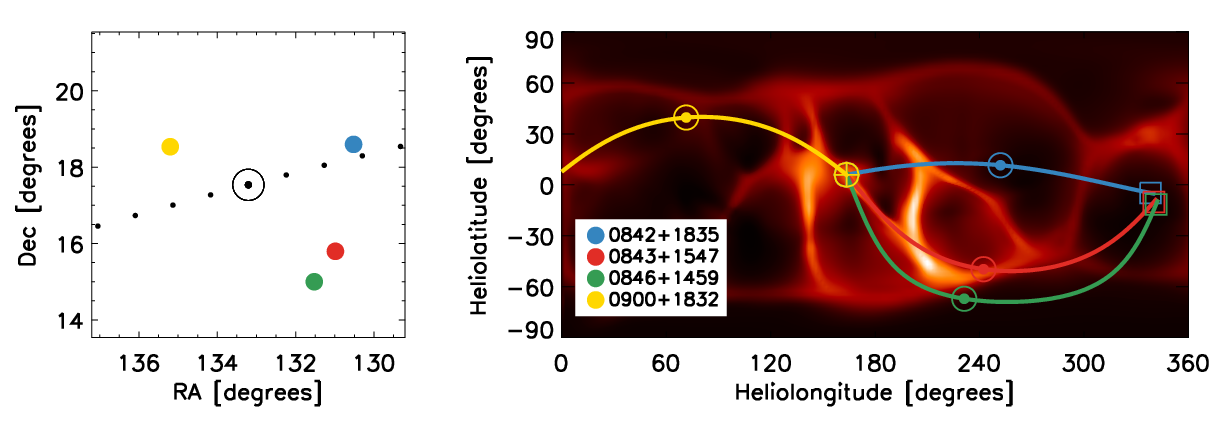}
\includegraphics[width=4.3cm]{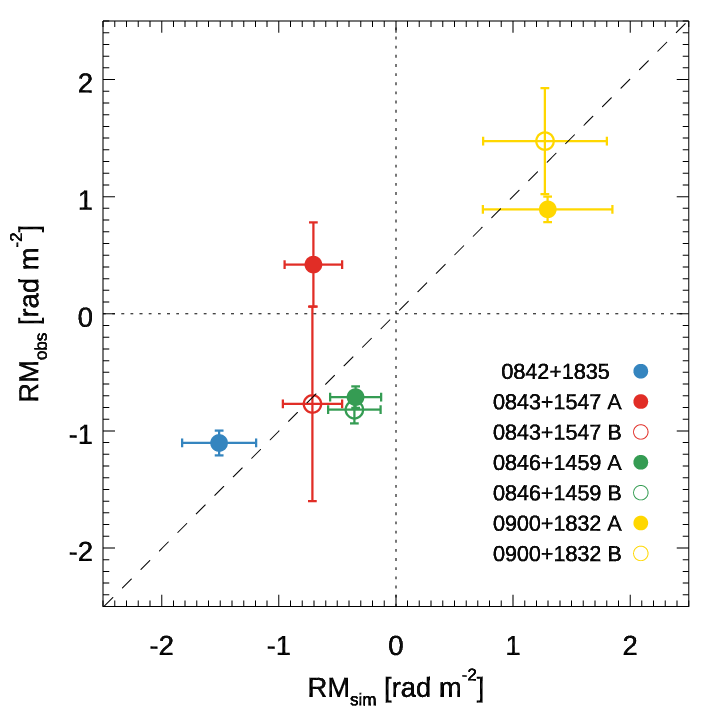}
\caption{
{\it Left panel}: 
Location of the Sun on August 2, 2012, at 16:00 UT, as indicated by the bull's-eye. The colored points mark the positions of the radio sources. The dotted line marks the ecliptic.
{\it Middle panel}: 
The synoptic map of the electron density is shown at 10.8 \rsun\ for CR 2126, as calculated by the MAS model. The curved lines represent the ray paths of the Faraday RM from four radio sources at 16:00 UT, projected onto their points of closest approach to the Sun. The bull's-eyes mark the closest approach points; the crossed circle indicates the position of Earth; and the square represents the projected location of the radio source. 
{\it Right panel}: 
Comparison of model-predicted and observed RMs. The dashed line represents perfect agreement. The product of the MHD MAS model’s electron densities and magnetic field intensities was scaled by a factor of 1.15 to match the Faraday rotation observations performed on August 2, 2012.
}  
\label{Fig5}
\end{figure*}

\section{The background corona}

To characterize the coronal environment in which the CME and the shock propagated, we employed an MHD modeling approach, enhanced by Faraday rotation observations to calibrate the model.

\subsection{MHD model}

The global coronal structure at the time of the eruption was modeled using the thermodynamic mode of the Magnetohydrodynamic Algorithm outside a Sphere (MAS) code (\citealt{lionello2009}) developed by Predictive Science Inc.\footnote{http://www.predsci.com}
The MAS model is a 3D, time-dependent, and finite-difference model that solves the Maxwell equations and the continuity, momentum, and energy equations. 
This MHD model requires photospheric radial magnetic field data and a heating mechanism as input conditions. 
To set the inner boundary conditions, the model uses magnetic field data from photospheric magnetograms obtained by the Heliospheric and Magnetic Imager (HMI; \citealt{scherrer2012}) on board the SDO (\citealt{pesnell2012}). 
The model incorporates detailed thermodynamics, including energy equations that account for thermal conduction along the magnetic field, radiative losses, and parameterized coronal heating (\citealt{lionello2001, lionello2009}).
The model's outer boundary is set at 30 \rsun. 
The MAS model is able to reproduce large-scale coronal features observed in white-light, EUV, and X-ray wavelengths and, as a result, effectively simulates the global coronal structure, including the location of streamers.

For this study, we used MAS model outputs that include global electron densities, magnetic field strengths, sound speeds, and solar wind speeds. 
These quantities are essential to compute the fast-mode Mach number for the eruption. 
However, it is important to note that the electron density ($n_e$) and the magnetic field strength ($B$) in the model rarely match the empirical values. 
Consequently, derived quantities, such as the Alfv\'en speed and the fast-mode Mach number, may not be entirely reliable.
In fact, both $B$ and $n_e$ must be scaled using correction factors ($f_b$ and $f_n$, respectively) to account for the changing conditions in the corona over time and during different phases of the solar cycle (e.g., \citealt{rouillard2016, kouloumvakos2019}).

\begin{table}[b]
\centering
\caption{Faraday rotation observations obtained on August 2, 2012, from \cite{kooi2017}.}
\begin{tabular}{c c c c}
\hline\hline
Source name  & RA (J2000)  & Dec (J2000) & Distance \\
             & [h \hspace{0.1cm} m \hspace{0.1cm}  s] & [$ ^\circ \hspace{0.2cm} \arcmin \hspace{0.2cm} \arcsec]       $  & [\rsun] \\
\hline
0842+1835   & 08 42 95.1 & +18 35 41 & 9.6 - 10.6\\
0843+1547 A & 08 43 56.5 & +15 47 41 & 9.9 - 10.5\\
0843+1547 B & 08 43 56.3 & +15 47 49 & 9.9 - 10.5\\
0846+1459 A & 08 46 05.9 & +14 59 54 & 11.1 - 11.4\\
0846+1459 B & 08 46 04.0 & +14 58 57 & 11.1 - 11.4\\
0900+1832 A & 09 00 48.4 & +18 32 01 & 8.0 - 8.6\\
0900+1832 B & 09 00 48.2 & +18 32 43 & 8.0 - 8.6\\
\hline\hline
\end{tabular}
\label{tab:Tab01}
\end{table}

\subsection{Faraday rotation observations}

As discussed above, Faraday rotation observations of extragalactic radio sources occulted by the solar corona offer a unique advantage in enabling a form of coronal tomography.
These observations allow us to simultaneously calculate the RM along various LOSs and provide an empirical estimate of the product of the longitudinal component of the magnetic field ($B_\parallel$) and the electron density ($n_e$) along each one.
\citealt{kooi2017} recently reported Faraday rotation observations performed on August 2, 2012, during the annual solar occultation of extragalactic radio sources 0842+1835, 0843+1547, 0846+1459, and 0900+1832. 
These observations were conducted at GHz frequencies using the upgraded Karl G. Jansky VLA of the NRAO and corrected for the negligible ionospheric Faraday rotation contribution. 
The data, with a resolution of approximately half an hour between each scan, are listed in Table 1 along with the heliocentric distances of the closest approach for each radio ray to the Sun.
While three of the four radio sources were affected by a CME passage (as detailed in \citealt{kooi2017}), all RM values reported are averages obtained prior to this event in the undisturbed corona. 
The left panel of Fig.~\ref{Fig5} shows the positions of the Sun and the four radio sources at 16:00 UT on August 2, 2012. 
The middle panel of the same figure shows the synoptic map of the electron density ($n_e$) at 10.8 \rsun\ for CR 2126, as calculated by the MAS model. 
The curved lines represent the ray paths of the RMs from the four radio sources, projected to their point of closest approach to the Sun. 

To calibrate the MHD model, we varied the correction factors $f_b$ and $f_n$ to achieve the best match between the modeled and observed RMs. 
The $f_b$ correction factor was applied to all components of the magnetic field to preserve the global topology. 
The right panel of Fig.~\ref{Fig5} shows the comparison between the modeled and observed RMs. 
The dashed line represents perfect agreement. 
By fitting the RM data using a weighted least-squares method (with a slope of $b = 1$ and intercept $a = 0$), we obtained a value for the product of the correction factors: $f_n \times f_b = 1.15$. 
In the next section, we discuss how these correction factors were determined based on the observed standoff distance reconstructed in Sect. 2.

\section{Results and discussion}

When the 3D morphology of a CME and its associated shock is accurately known, white-light observations of the CME-driven shock can be used to probe the physical properties of the coronal plasma through which the eruption propagates (e.g., \citealt{mancuso2019} and references therein). 
The shape, size, and standoff distance of a CME-driven shock are influenced by several factors, including the CME's own shape and size, its velocity relative to the local plasma and fast-mode magnetosonic speed, and key properties of the medium, such as the ratio of specific heats ($\gamma$), magnetic field strength, and orientation.

Coronal-mass-ejection-driven shocks can generally be classified into two types: bow shocks and expansion (piston-driven) shocks (\citealt{vrsnak2008}; \citealt{warmuth2015}). 
Bow shocks form when solar wind flows around relatively narrow CMEs, while expansion shocks arise ahead of a piston driver that expands outward through and accumulates solar wind (\citealt{zic2008}; \citealt{lulic2013}). 
The plasma is pushed outward by the expanding piston with no material flowing behind. 
Once the piston decelerates, the shock detaches and continues to propagate. 
Although the distinction between piston-driven and bow shocks can be unclear, observational data suggest that fast CMEs with widths $\lesssim 60^\circ$ are typically bow shocks, while those with widths $> 60^\circ$ are primarily piston-driven (\citealt{kahler2019}; \citealt{fainshtein2019}). 
Generally, both types of shocks are likely to occur simultaneously.
In fact, rapid expansion in the early phase can generate a piston-driven shock, while continuous radial motion of the CME ejecta could produce a bow shock (\citealt{kwon2014}).

Shocks propagate at speeds $v_{\rm sh}$ that exceed the characteristic speed of the upstream medium. 
In a magnetized plasma, there are typically three modes: fast magnetosonic waves, slow magnetosonic waves, and intermediate Alfv\'en waves. 
However, CME-associated shocks are usually interpreted as collisionless fast-mode shock waves. 
The fast-mode speed, $v_{\rm fm}$ is given by
\begin{equation}
    v_{\rm fm} = \sqrt{{1\over2} \left[v_{\rm A}^2 + c_{\rm S}^2 + \sqrt{(v_{\rm A}^2 + c_{\rm S}^2)^2 - 4v_{\rm A}^2c_{\rm S}^2\cos^2{\theta_{\rm Bn}}}\right]},
\end{equation}
where $v_{\rm A}$ is the Alfvén speed, $c_{\rm S}$ is the sound speed, and $\theta_{\rm Bn}$ is the angle between the wave vector and the magnetic field vector. 
This fast-mode speed is the characteristic speed we use to compare with the shock front speed. 
The fast-mode Mach number is defined as $M_{\rm fm} = (v_{\rm sh}-v_{\rm sw})/v_{\rm fm}$, where $v_{\rm sh}$ is the shock speed in the observer frame.
For parallel propagation ($\cos{\theta_{\rm Bn}} = 1$), the fast-mode speed simplifies to
\begin{equation}
   v_{\rm fm,\parallel} = \sqrt{{1\over2} (v_{\rm A}^2 + c_{\rm S}^2 + |v_{\rm A}^2 - c_{\rm S}^2|)}.
\end{equation}
For propagation across magnetic field directions, 
\begin{equation}
   v_{\rm fm,\perp} = \sqrt{v_{\rm A}^2 + c_{\rm S}^2},
\end{equation}
and at intermediate angles, the fast-mode wave speed lies between these two extremes.

\begin{figure*}
\centering
\includegraphics[width=14cm]{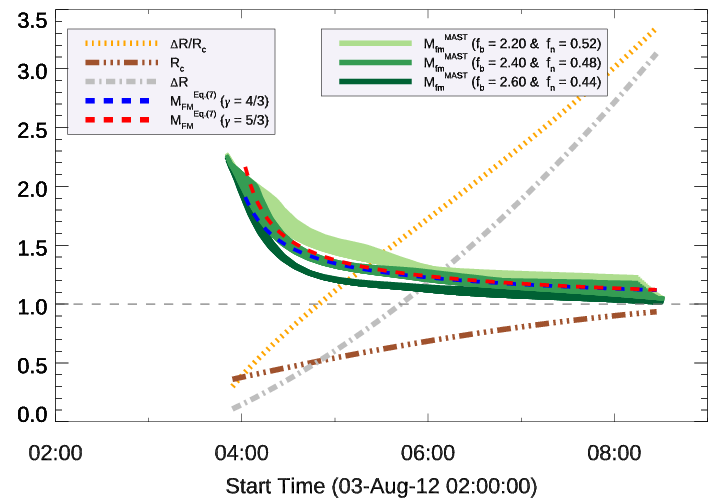}
\caption{
Fast-mode Mach number $M_{\rm fm}$ as a function of time calculated using Eq. (7) for $\gamma = 4/3$ ({\it dashed blue curve}) and $\gamma = 5/3$ ({\it dashed red curve}). 
The {\it shaded green bands} represent $M_{\rm fm}$ values computed using MAS with the given correction factors, as the angle $\theta_{\rm Bn}$ between the wave vector and the magnetic field vector varies from 0\dg to 25\dg. 
The plot also shows the temporal evolution of the shock standoff distance $\Delta R$, the effective radius of the flux rope's curvature $R_c$, and the ratio $\Delta R/R_c$.
}  
\label{Fig6}
\end{figure*}

The standoff distance between the shock and the CME's leading edge, $\Delta R = r_{\rm sh} - r_{\rm cme}$, depends on the CME's shape and the relative Mach number of the flow. 
Early gasdynamic studies with spherical obstacles found that $\Delta R$ is linearly related to the inverse of the compression ratio $X = \rho_d / \rho_u$, where $\rho_u$ and $\rho_d$ represent the densities upstream and downstream, respectively (\citealt{seiff1962}). 
However, this relationship is more complex for nonspherical structures such as those CMEs modeled as flux ropes. 
In a hydrodynamic regime based on mass conservation, the compression ratio $X$ can be expressed in terms of the upstream Mach number $M_u$ and the ratio of specific heats $\gamma$ (\citealt{priest2012}):
\begin{equation}
    {\rho_u\over \rho_d} = {(\gamma - 1)M_u^2 + 2\over (\gamma + 1)M_u^2}.
\end{equation}
However, this equation is only applicable when the shock velocity and magnetic field are parallel. 
In an MHD regime, the assumption of quasi-parallelity is reasonable for shock propagation at higher altitudes in the corona, where the magnetic field is predominantly radial.
Thus, for CME-driven shocks, the hydrodynamic case provides a plausible approach.
For low Mach numbers ($M_u \lesssim 3$), \cite{Farris1994} derived the following expression for $\Delta R$:
\begin{equation}
    {\Delta R\over R_{\rm c}} = 0.81 {(\gamma - 1)M_u^2 + 2\over (\gamma + 1)(M_u^2 - 1)},
\end{equation}
where $R_{\rm c}$ is the CME's radius of curvature. 
In the astrophysical context, hydrodynamic relations offer a good approximation for low $M_{\rm fm}$ (\citealt{russell2002}).

For a CME modeled as a flux rope with a croissant-like shape, the standoff distance is governed by the geometric mean of the two radii of curvature perpendicular to the direction of propagation (\citealt{stahara1989}; \citealt{russell2002}; \citealt{kay2020}). 
The minor and major radii of curvature are provided by the azimuthal extent of the rope (perpendicular to the solar wind flow and the rope axis), and the major axis is provided by the curvature of the axial field of the rope (\citealt{russell2002b}). 
The minor and major radii can be inferred from the coronagraph images. 
For this event, the difference images from COR2-A provide a clear face-on view of the flux rope, so that we were able to retrieve the minor and major radii of curvature as they evolved self-similarly with time.
Given the quantities $\Delta R$ and $R_{\rm c}$ deduced from the coronagraphic images, and assuming a reasonable value for $\gamma$, the fast-mode Mach number $M_{\rm fm}$ can be easily obtained by inverting Eq. (6):
\begin{equation}
    M_{\rm fm} = \sqrt{{2 \times 0.81 + {\Delta R\over R_{\rm c}}(\gamma + 1)}\over{\Delta R\over R_{\rm c}}(\gamma + 1) -0.81(\gamma - 1)}.
\end{equation}

Figure~\ref{Fig6} shows the standoff distance $\Delta R$ of the shock derived from the 3D reconstruction as a function of time. 
The figure also shows the temporal dependence of the effective radius of curvature of the flux rope, $R_{\rm c}$, and of the ratio $\Delta R/R_{\rm c}$. 
These parameters clearly increase quasi-linearly over time. 
The same figure further illustrates the time evolution of the fast-mode Mach number, calculated using Eq. (7) for the two common cases of adiabatic indices $\gamma = 4/3$ and $5/3$.
In the above analysis, we assumed quasi-parallel propagation above 6 \rsun, where the angle $\theta_{Bn}$ between the shock wave vector and the magnetic field vector was varied between 0\dg\ and 25\dg. 
To derive our result, we also assumed that the shock formed as the perturbation propagated through the densest region of the corona. 
This region approximately corresponds within a few degrees to the nose of the shock front.
We also averaged the data over CR longitudes between 78\dg\ and 90\dg\, with a mean value of 84\dg, inferred from the shock reconstruction.
Adjusting the correction factors, $f_n$ and $f_b$--constrained by the previously determined product, $f_n \times f_b = 1.15$, derived from the Faraday RMs--and comparing the result with that obtained from Eq. (7), we find that $M_{\rm fm}$ from the MAS model is best described with $f_n \simeq 2.4$ and $f_b \simeq 0.5$, with $\gamma = 5/3$ (see Fig.~\ref{Fig6}). 
Our results, deduced from Fig.~\ref{Fig6}, indicate that the fast-mode Mach number decreased from about 2 to nearly 1 over a period of 4 hours. 
This suggests that the shock gradually dissipated and could not propagate above approximately 20 \rsun.
Moreover, the low values we obtained for the fast-mode Mach number appear to explain the nondetection of type II radio emission both in the extended corona and in interplanetary space.
Type II radio bursts are characterized as emission bands that slowly drift from high to low frequencies in radio dynamic spectra, with a duration of several minutes, and are usually interpreted as signatures of CME-driven shocks (e.g., \citealt{Mancuso2002, mancuso2008,bemporad2010}).
The Mach number was presumably subcritical (e.g., \citealt{edmiston1984,bemporad2011}), and therefore implies that the particle acceleration efficiency was likely very low (\citealt{mann2003}). 
Consequently, the weak shock might not accelerate a sufficient number of electrons required to produce a type II burst (see \citealt{gopalswamy2008,gopalswamy2010b} for a discussion about radio-quiet CME-driven shocks).

This outcome is consistent with \cite{kouloumvakos2019}, who also applied estimated correction factors to scale up the coronal magnetic field and scale down the density values provided by the MHD models.
These correction factors were obtained by extrapolating the averaged values of the unsigned radial component of the magnetic field and solar wind density at an outer boundary within the MHD model together with an inverse square dependence. 
These correction factors were then compared with in situ measurements at 1 AU. 
According to \cite{kouloumvakos2019}, the magnetic field scaling factor, $f_b$, ranges between 1.6 and 2.4, averaging at 1.8, while the density correction factor, $f_n$, ranges between 0.3 and 0.6, averaging at 0.45. 
Our results show that the MAS model similarly overestimates densities and underestimates magnetic fields near the Sun, particularly at heliocentric distances greater than 8 \rsun\ (the shortest proximity along the LOS; see Table 1). 

\section{Summary and conclusion}

In this work, we introduced a novel approach that combined advanced MHD modeling with Faraday rotation observations to improve our understanding of CMEs, their associated shocks, and their impact on space weather. 
Faraday rotation measurements of extragalactic radio sources occulted by the solar corona provide a powerful complementary tool for probing the pre-eruption electron density and magnetic field structure of the solar corona and thus improve predictions from global MHD models. 
In fact, these observations offer independent information about the integrated product of the LOS magnetic field component and the electron density.
Using the August 3, 2012, CME as a case study, we investigated the 3D geometry and kinematics of the shock driven by a narrow flux rope. 
Observations were obtained from three spacecraft: SOHO, STEREO-A, and STEREO-B. 
Strategically positioned at the time of the event, these instruments allowed us to capture the shock's propagation in space.

The fortunate availability of Faraday RMs, obtained a few hours before the transient, allowed us to properly calibrate the MHD MAS model in its thermodynamic mode (\citealt{lionello2009}). 
To perform this calibration, we adjusted for the plasma conditions using correction factors for the magnetic field ($f_b$) and electron density ($f_n$). 
These were derived from Faraday rotation data. 
The application of these factors ($f_b \simeq 2.4$ and $f_n \simeq 0.5$) produced a strong match between the fast-mode Mach number predicted by MAS and that obtained from the analysis of the 3D reconstruction of coronagraphic data.
The reconstruction utilized the model from \cite{Farris1994} and \cite{russell2002}. 
This result is consistent with \cite{kouloumvakos2019}, who also applied estimated correction factors to scale up the coronal magnetic field and to scale down the density values provided by the MHD models.
In fact, the Mach number was probably subcritical, and we expected the particle acceleration efficiency to be particularly low.
Our analysis shows that the shock's fast-mode Mach number decreased over time, approaching unity as the shock dissipated. This is consistent with the expected physical behavior of shocks as they lose energy during propagation.

In conclusion, this study underscores the critical role of integrating observational techniques, such as Faraday rotation and MHD modeling, to improve the understanding of CMEs and their impact on space weather. 
The results of our analysis contribute to a more reliable framework for predicting space weather events, particularly the timing and intensity of geomagnetic storms caused by solar eruptions. 
Further refinements in these models, alongside continued observational efforts, are vital to enhance space weather forecasting capabilities and mitigate the risks to space-based technologies and the Earth’s magnetosphere caused by solar events.\\

\textbf{Acknowledgements:}
The author thanks the anonymous referee for a thorough review, providing constructive comments and suggestions.
The SECCHI data are produced by an international consortium of the NRL, LMSAL, and NASA GSFC (USA), RAL and U. Bham (UK), MPS (Germany), CSL (Belgium), and IOTA and IAS (France). 
The SOHO/LASCO data used here are produced by a consortium of the Naval Research Laboratory (USA), Max-Planck-Institut für Aeronomie (Germany), Laboratoire d’Astronomie (France), and the University of Birmingham (UK). 
SOHO is a project of international cooperation between ESA and NASA. 
STEREO is a mission in NASA’s Solar Terrestrial Probes program. 
The LASCO CME catalog is generated and maintained at the CDAW Data Center by NASA and The Catholic University of America in cooperation with the Naval Research Laboratory. 
We also acknowledge Predictive Science Inc. for publicly providing the data used in this study. 
This study has used the JHelioviewer software provided by ESA. 
This research has used PyThea v0.12.0, an open source and free Python package to reconstruct the 3D structure of CMEs and shock waves (Zenodo: https://doi.org/10.5281/zenodo.5713659).
Finally, the author wishes to express his sincere gratitude to Prof. Steven R. Spangler for being a source of great inspiration and wisdom.

\bibliographystyle{aasjournal}
\bibliography{biblio.bib}

\end{document}